\documentclass{aa}
\usepackage{graphicx}
\usepackage{txfonts}
\usepackage{mathrsfs}

 \def\mso{\,\mathrm{M}_\odot}
 
 \def\lso{\,{\rm L}_\odot}

 \def\simle{\mathrel{\hbox{\rlap{\hbox{\lower4pt\hbox{$\sim$}}}\hbox{$<$}}}}
 \def\simgr{\mathrel{\hbox{\rlap{\hbox{\lower4pt\hbox{$\sim$}}}\hbox{$>$}}}}

 \def\c2{^{12}{\mathrm C}}
 \def\c3{^{13}{\mathrm C}}
 \def\n14{^{14}{\mathrm N}}
 \def\c1213{^{12}{\mathrm C}/^{13}{\mathrm C}}
 \def\he3he4{^3\mathrm{He}/^4\mathrm{He}}
 \def\he3{^3\mathrm{He}}
 
 \def\Te{T_{\rm eff}}
 \def\Ge{\Gamma_{\rm e}}

\begin{document}
   \title{The spectroscopic Hertzsprung-Russell diagram}

   \author{N. Langer\inst{1}
           \and
          R.P. Kudritzki\inst{2,3}}

   \institute{Argelander-Institut f\"ur Astronomie der Universit\"at Bonn, Auf dem H\"ugel 71, 53121 Bonn, Germany
  \and
   Institute for Astronomy, University of Hawaii, 2680 Woodlawn Drive, Honolulu, HI 96822, USA
  \and
   University Observatory Munich, Scheinerstr. 1, D-81679 Munich, Germany}

   \date{2014 / 2014}

  \abstract
  {The Hertzsprung-Russell diagram is an essential diagnostic diagram for
stellar structure and evolution, which has now been
in use for more than 100 years. 
}
  {We introduce a new diagram based on the gravity-effective temperature
diagram, which has various advantages.  }
  {Our spectroscopic Hertzsprung-Russell (sHR) diagram shows the inverse of the 
flux-mean gravity versus the effective temperature.
Observed stars whose spectra have been quantitatively analyzed can be entered in this diagram
without the knowledge of the stellar distance or absolute brightness.
}
{Observed stars can be as conveniently compared to stellar evolution calculations
in the sHR diagram as in the Hertzsprung-Russell diagram. 
However, at the same time, our ordinate is proportional to the stellar mass-to-luminosity
ratio, which can thus be directly determined. For intermediate- and low-mass star
evolution at constant mass, we show that the shape of an evolutionary track in the
sHR diagram is identical to that in the Hertzsprung-Russell diagram.
We also demonstrate that for hot stars, their stellar Eddington factor 
can be directly read off the sHR diagram. For stars near their Eddington limit,
we argue that a version of the sHR diagram may be useful where the gravity is exchanged 
by the effective gravity.
} 
  {We discuss the advantages and limitations of the sHR diagram, and show that it can be
fruitfully applied to Galactic stars, but also to stars with known distance, 
e.g., in the LMC or in galaxies beyond the Local Group.
}


   \maketitle


\section{Introduction}

The Hertzsprung-Russell (HR) diagram has been an important diagram for the 
understanding of stellar evolution
for more than a hundred years (Nielsen 1964). Hertzsprung (1905) and later independently Russell (1919)
realized that the knowledge of the absolute brightness of stars 
together with their spectral type or surface temperature allowed
fundamentally different types of stars to be distinguished.  

Hertzsprung and Russell had already realized that the apparent stellar brightness was insufficient to draw
conclusions, but that the absolute brightnesses, i.e., the distances, are required to 
properly order the stars in the HR diagram. 
Order can also be achieved for stars in star clusters where the distance may still be unknown, 
but the distances of all stars are roughly equal, in what we now call color-magnitude diagrams 
because the apparent and absolute brightness differences are equal.

Hertzsprung and Russell pointed out that the HR diagram contains information about
the stellar radii, with the giant sequence to be found at a larger brightness but similar surface temperatures
to the cools stars of the dwarf or main sequence. And indeed, it remains one of the main advantages of the
quantitative HR diagram that stellar radii can be immediately determined, thanks to the Stefan-Boltzmann law.

Later, with the advent of stellar model atmosphere calculations, 
it became possible to quantitatively derive accurate stellar
surface gravities (see, e.g., Auer and Mihalas 1972 and references therein).
This allowed stars to be ordered in the 
surface gravity-effective temperature ($g-T_{\rm eff}$) diagram 
(sometimes called the Kiel diagram), since a larger surface gravity for stars of a given surface temperature
can imply a larger mass (Newell 1973, Greenstein and Sargent 1974). 
The main advantage of the $g-T_{\rm eff}$ diagram is that stars can be compared to stellar
evolution predictions  without the prior knowledge of their distance
(a first  example is given in Kudritzki 1976). However,
the radius or any other stellar property can not be directly
identified from the $g-T_{\rm eff}$ diagram. 
Moreover, the comparison with stellar evolution calculations is often negatively affected by the relatively large 
uncertainties of the spectroscopic gravity determinations.

In this paper, we want to introduce a diagnostic diagram for stellar evolution which combines the advantages 
of the Hertzsprung-Russell and of the $g-T_{\rm eff}$  diagram. We introduce the 
spectroscopic Hertzsprung-Russell (sHR) diagram in Sect.~2 and
compare it with the Hertzsprung-Russell and the $g-T_{\rm eff}$ diagram in Sect.~3.
We discuss stars with changing mass and helium abundance in Sect.~4, 
and focus on stars near the Eddington limit in Sect.~5, and on low- and intermediate-mass stars in
Sect.~6. Finally, we give an example for the application of the sHR diagram in Sect.~7,
and close with concluding remarks in Sect.~8.

\section{The sHR diagram}

The idea of the sHR diagram is to stick to the variables surface gravity and effective temperature,
as those can be directly derived from stellar spectra without knowledge of the stellar distance.
We then define the quantity ${\mathscr L} := T_{\rm eff}^4/g$, which is the inverse of the
``flux-weighted gravity'' defined by Kudritzki et al. (2003). Kudritzki et al. (2003, 2008)
showed that $1/{\mathscr L}$ --- and thus ${\mathscr L}$ as well --- is expected to remain almost
constant in massive stars, because combining 
\begin{equation}
g = {G M \over R^2} ,
\end{equation}
where $M$ and $R$ are stellar mass and radius, $G$ the gravitational constant, and $g$ the stellar
surface gravity, with the Stefan-Boltzmann law
\begin{equation}
L = 4\pi\sigma R^2\Te^4 ,
\end{equation}
with $L$ and $\Te$ representing the stellar bolometric luminosity and effective temperature, and $\sigma$
being the Stefan-Boltzmann constant, one finds that
\begin{equation}
L = 4\pi\sigma G M {\Te^4\over g} = 4\pi\sigma G M{\mathscr L} .
\end{equation} 
\begin{figure*}[]
    \centering
     \includegraphics[angle=-90,width=8.5cm]{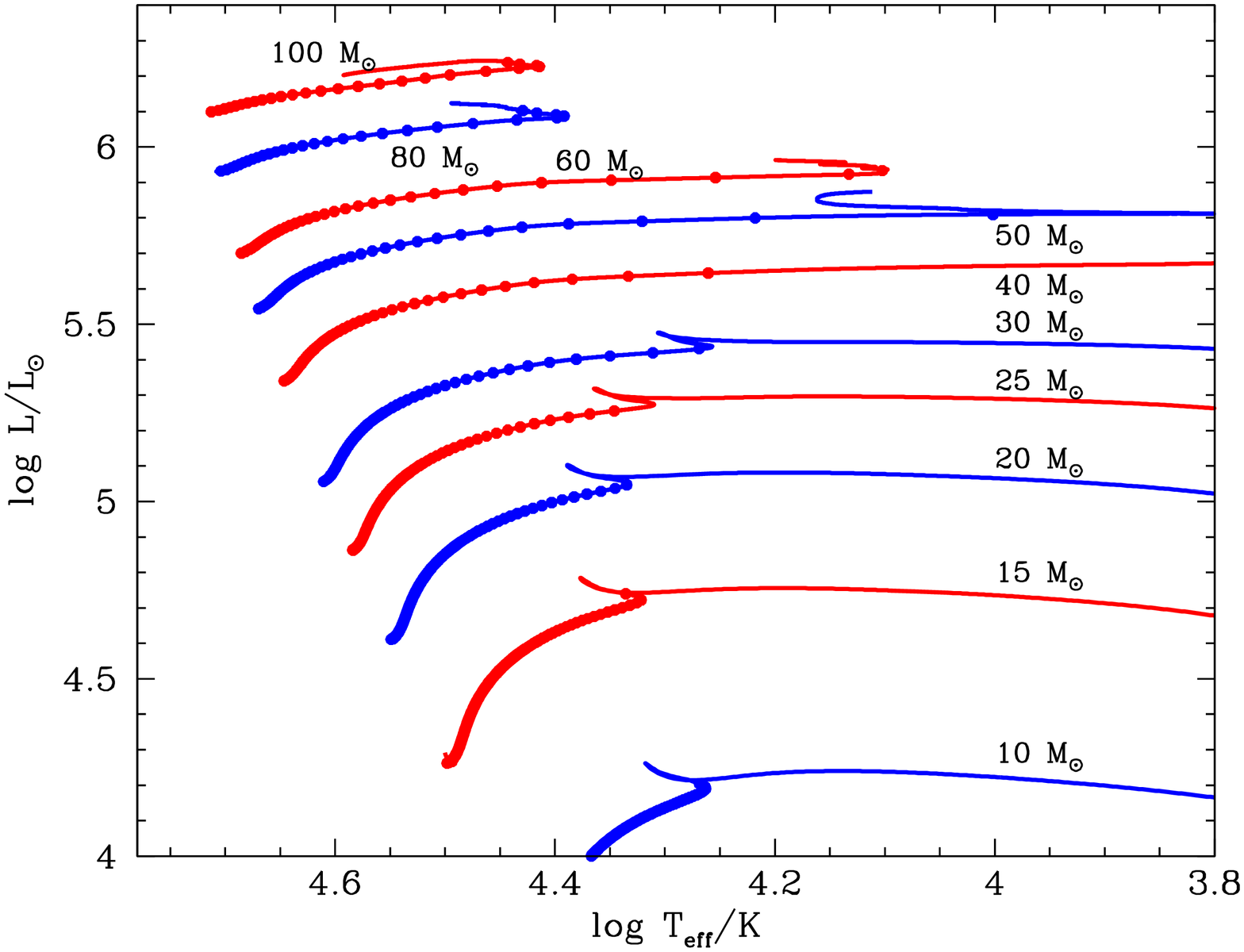}\includegraphics[angle=-90,width=8.5cm]{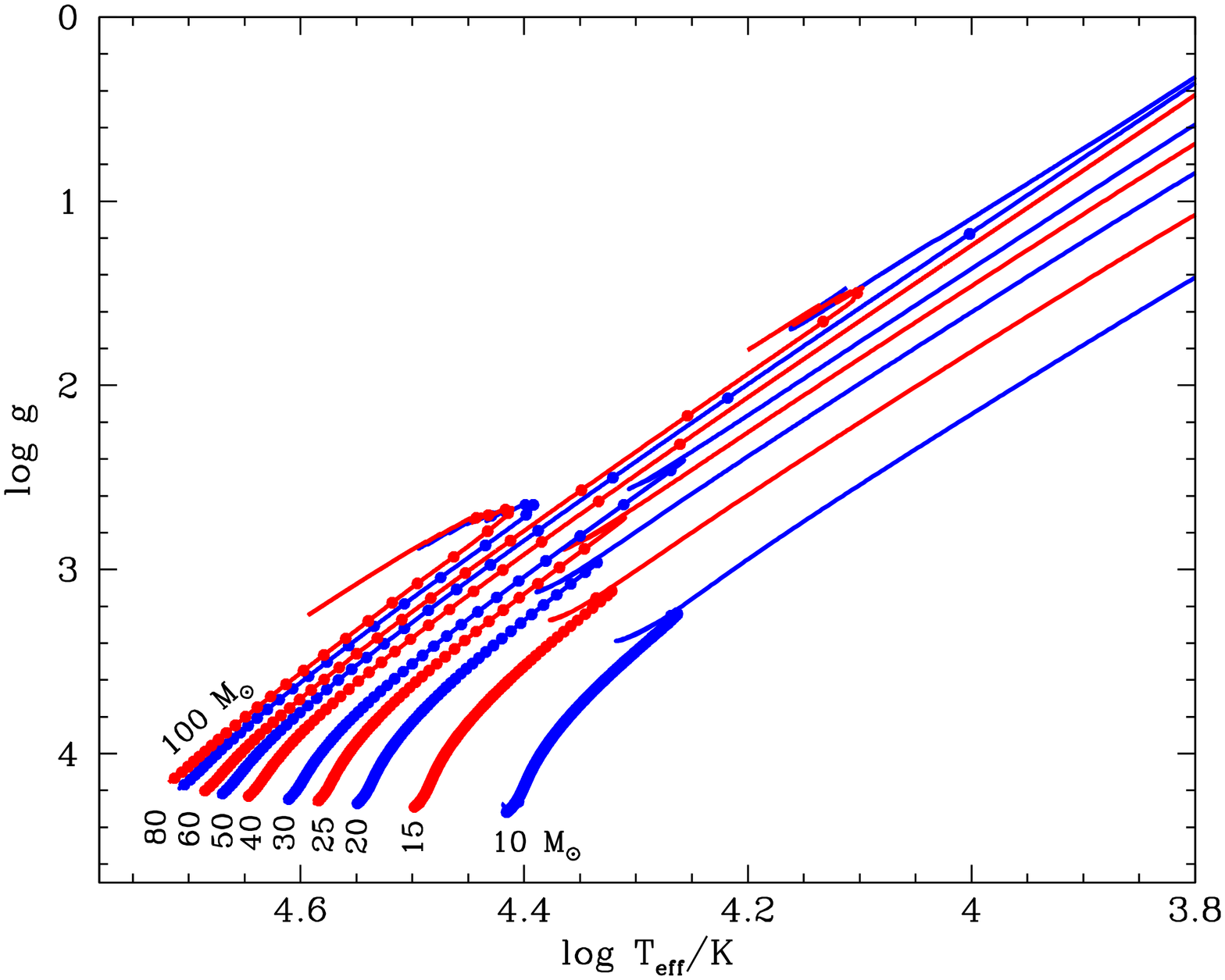}
    \caption{Evolutionary tracks of stars initially rotating with an equatorial velocity of
$\sim 100\,$km/s, with initial masses in the range $10\mso\dots 100\mso$, in the HR diagram (left),
and in the $g-T_{\rm eff}$ diagram (right).
The initial composition of the models
is solar. The models up to $60\mso$ are published by Brott et al. (2011), while those of higher mass
are unpublished additions with identical input physics.}
    \label{fig_grid}
\end{figure*}
The fact that massive stars evolve at nearly constant luminosity (see below) allowed Kudritzki et al. (2003, 2008)
to use the flux-weighted gravity-luminosity relationship as a new method for deriving extragalactic distances.
Since for stars of constant mass, 
${\mathscr L} \sim L $, ${\mathscr L}$ behaves like the stellar
luminosity, and the evolutionary tracks in the ${\mathscr L}- T_{\rm eff}$ diagram
partly resemble those in the Hertzsprung-Russell diagram. In a sense, the sHR diagram is a version
of the $g-T_{\rm eff}$ diagram where the stellar evolutionary tracks of massive stars are horizontal again,
which allows for a better visual comparison (see Sect.~3). 
More importantly, since for hot massive stars the spectroscopic determination of the 
flux weighted gravity $g_{\rm F} = g/T_{\rm eff}^4$ is less
affected by the uncertainties of temperature than the determination 
of gravity (see sect.\,6.1 in Kudritzki et al. 2012, for a detailed
physical explanation), $g_{\rm F}$can be determined more precisely than $g$. 
A good example is the case of blue supergiant stars, where the
uncertainty of $\log g_{\rm F}$ is about 0.05\,dex, while the error in $\log g$ is 
two or three times as large (see Kudritzki et al. 2008 and
2012, Tables 1 and 2, respectively).

However, the sHR diagram is not just a rectified version of the $g-T_{\rm eff}$ diagram, or a distance independent
version of the Hertzsprung-Russell diagram. Its deeper meaning becomes obvious when we write 
Eq.~(3) as
\begin{equation}
{\mathscr L} = {1\over 4\pi\sigma G} {L\over M} ,
\label{eq_lm}
\end{equation}
or, with the Eddington factor $\Gamma = L/L_{\rm Edd}$ and $L_{\rm Edd}=4\pi c G M/\kappa$ as
\begin{equation}
{\mathscr L} = {c \over \sigma\kappa} \Gamma ,
\label{eq_ga}
\end{equation}
where $c$ is the speed of light and $\kappa$ the radiative opacity at the stellar surface.
Obviously, ${\mathscr L}$, for a given surface opacity, is directly proportional to the 
luminosity-to-mass ratio (Eq.\,\ref{eq_lm}) and to the
stellar Eddington factor  $\Gamma$ (Eq.\,\ref{eq_ga}).
 
Again, this is particularly useful for massive stars, where the Eddington factor is not extremely small anymore  
and can approach values close to unity.
In addition, in hot massive stars, the radiative opacity is dominated by electron scattering (Kippenhahn \& Weigert, 1990),
which can be approximated as
\begin{equation}
\kappa \simeq \kappa_{\rm e}= \sigma_{\rm e} (1+X) ,
\end{equation}
with the cross section for Thomson scattering $\sigma_{\rm e}$ and the surface hydrogen mass fraction $X$.
Consequently, for massive stars with unchanged surface abundances, ${\mathscr L}$ is truly proportional
to the Eddington factor. Furthermore, for helium enriched stars, the helium abundance can also be determined
from model atmosphere analyses, and the opacity can be corrected accordingly. 

The near proportionality of ${\mathscr L}$ to the Eddington factor implies a fundamental difference
between the sHR and the HR or the $g-T_{\rm eff}$ diagrams. In contrast to the last two, the sHR diagram
has an impenetrable upper limit, i.e.,  the Eddington limit. For example, since for large mass
the mass-luminosity exponent $\alpha$ in the mass luminosity relation $L\sim M^{\alpha}$
tends asymptotically to $\alpha=1$ (Kippenhahn \& Weigert 1990), even stars of extremely high mass
may not violate the Eddington limit, which means that there is no upper bound on the luminosity
of stars in the HR diagram. In contrast, in the sHR diagram for hot stars of normal composition ($X \simeq 0.73$),
we find that $\log {\mathscr L} / {\mathscr L}_{\odot} \simeq 4.6$ is the maximum achievable value.

\subsection{Limitations of the sHR diagram}

The sHR diagram introduced above can be produced from any stellar evolution models without
limitations, and any star for which effective temperature and surface gravity are measured can be entered
in this diagram. However, the interpretation of a comparison of observed stars with stellar models in this
diagram has some restrictions. 

In order to be able to interpret the ordinate of the sHR diagram in terms of an Eddington factor,
the opacity that applies to the surfaces of the stars (modeled or observed) needs to be the same
for all stars considered in the diagram. As mentioned above, this is approximately given for hot stars
with the same hydrogen abundance, and thus holds for most Galactic OB stars.

For helium-enriched hot stars, their Eddington factor can still be read off the sHR diagram, as long as
the helium abundance is known (e.g., from a stellar atmosphere analysis) because the 
electron fraction can then be computed, and the electron scattering opacity can be computed from Eq.~6.

For cool stars, the situation becomes more complicated since hydrogen and helium may be partly recombined.
This becomes noticeable for hydrogen for $T\simle 10\,000\,$K and strong for $T\simle 8\,000\,$K.
For helium, the first electron recombines at $T\simeq 28\,000\,$K, while the recombination temperature for
the second electron is similar to that of hydrogen. This means that for stars with temperatures below $T\simeq 28\,000\,$K
the electron-scattering Eddington factor as determined from the sHR diagram may only be correct to within
$\sim$10\%.  Of course, a precise electron fraction can be obtained from stellar model atmosphere calculations.

\begin{figure*}[]
    \centering
     \includegraphics[angle=-90,width=12.5cm]{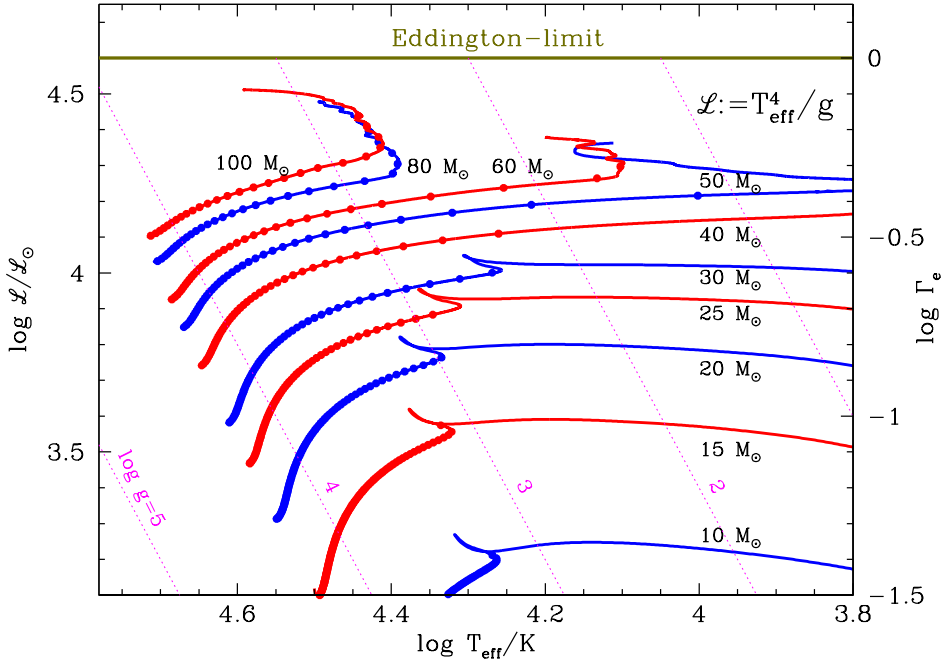}
    \caption{Tracks of the same models as those shown in Fig.~\ref{fig_grid}, in the sHR diagram.
While the ordinate is defined via the spectroscopically measurable quantities $\Te$ and $\log g$,
its numerical value gives the logarithm of the luminosity-to-mass ratio, in solar units.
The right-side ordinate scale gives the atmospheric Eddington factor for hot hydrogen-rich stars.
The horizontal line at $\log {\mathscr L} / {\mathscr L}_{\odot} \simeq 4.6$ indicates the
location of the Eddington limit. The dotted straight lines are lines of constant $\log g$, as indicated.  
}
    \label{fig_KiBo}
\end{figure*}

\section{Comparison of the diagrams}

Figure~\ref{fig_grid} shows contemporary evolutionary tracks for stars in the mass range $10\mso\dots 100\mso$ 
in the HR and in the $g-T_{\rm eff}$ diagram (cf., Langer 2012). 
They can be compared to tracks of the same models in the sHR diagram in Fig.~\ref{fig_KiBo}.
For the sHR diagram we plot the quantity ${\mathscr L} := T_{\rm eff}^4/g$
as function of the effective temperature of selected evolutionary sequences,
where $g$ is the surface gravity, and where ${\mathscr L}$ is normalized to
solar values for convenience (with $\log {\mathscr L}_{\odot} \simeq 10.61$).
With the above definition of the Eddington factor,
we obtain ${\mathscr L}  = c/(\kappa_e \sigma)\,\Gamma_e$, where $\sigma$ is the
Stefan-Boltzmann constant. For a given surface chemical composition,
${\mathscr L}$ is proportional to the Eddington factor $\Gamma_e$.
For Solar composition, we have
$\log {\mathscr L} / {\mathscr L}_{\odot} \simeq 4.6 + \log \Gamma_e$.

According to the definition of $\mathscr L$ in Sect.~2, lines of constant $\log g$
can be drawn as straight lines in the sHR diagram, as
\begin{equation}
4 \log T_{\rm eff} - \log {\mathscr L \over {\mathscr L}_{\odot}} = \log {\mathscr L}_{\odot} + \log g 
\end{equation}
(see Fig.~\ref{fig_KiBo}). We note that in the classical HR diagram, this is not possible, as two stars
with the same surface temperature and luminosity that have different masses occupy the same location
in the HR diagram, but since they must have the same radius, their surface gravities are different.
In the sHR diagram, both stars fall on different iso-$g$ lines.

\begin{table}[t]
\caption{Surface gravity (see Eq.\,(1)) of stars with a normal helium 
surface mass fraction ($Y\simeq 0.26$) near their Eddington limit, as
a function of their surface temperature, according to Eq.\,\ref{eq_g1}.}
\centering
\begin{tabular}{c c c c c c}
\hline\hline
$T_{\rm eff}$/kK =  & 100 & 50 & 40 & 30 & 20 \\
\hline
$\Ge \rightarrow 1$ & 4.82 & 3.61 & 3.22 & 2.72 & 2.02 \\
$\Ge = 0.9$ & 4.86 & 3.66 & 3.27 & 2.77 & 2.07 \\
$\Ge = 0.8$ & 4.91 & 3.71 & 3.32 & 2.82 & 2.12 \\
$\Ge = 0.7$ & 4.97 & 3.77 & 3.38 & 2.88 & 2.17 \\
$\Ge = 0.5$ & 5.12 & 3.91 & 3.52 & 3.03 & 2.32 \\
$\Ge = 0.3$ & 5.34 & 4.13 & 3.75 & 3.25 & 2.54 \\
$\Ge = 0.1$ & 5.81 & 4.61 & 4.22 & 3.72 & 3.02 \\
\hline
\end{tabular}
\end{table} 

Comparing the two diagrams in Fig.~\ref{fig_grid}, 
it becomes evident that the evolutionary tracks, especially of the very massive stars
in the $g-T_{\rm eff}$ diagram, are located very close together. For example, the tracks of the $80\mso$ and the $100\mso$ stars
can barely be distinguished. The reason is that  
\begin{equation}
g = 4\pi\sigma G \Te^4 {M\over L}  ,
\label{eq_g}
\end{equation}
and since the exponent $\alpha$ in the mass-luminosity relation tends to unity (see Sect.~2) 
we find that the gravities of very massive stars of the same effective temperature must become 
almost identical. Equation\ref{eq_g} determines the gravities of stars near the Eddington-limit,
as it transforms to
\begin{equation}
g = {\sigma \kappa_{\rm e} \over c} { \Te^4 \over \Ge}   
\label{eq_g1}
\end{equation}
(see Table~1). For stars with a helium-enriched surface, Eq.\,\ref{eq_g1} results in
smaller gravities because of the reduced electron scattering opacity.

\begin{figure*}[]
    \centering
     \includegraphics[angle=-90,width=13.5cm]{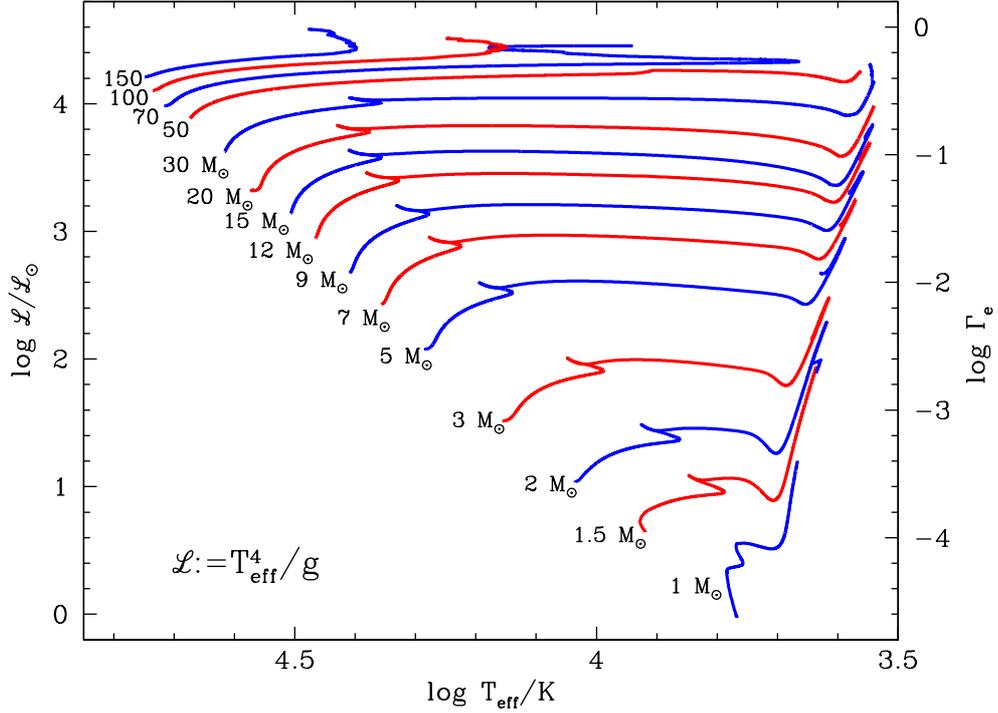}
   \caption{Combined sHR diagram for low-, intermediate- and high-mass stars. The tracks are computed using the
LMC initial composition and include those published in Brott et al. (2011) and K\"ohler et al. (2014).}
    \label{fig_all}
\end{figure*}

While this is a problem that remains true also for the sHR diagram,
Fig.\ref{fig_KiBo}  shows why it is somewhat remedied in this case. The gravities of stars
change by many orders of magnitude during their evolution, which is reflected in the
Y-axis of the $g-T_{\rm eff}$ diagram. Therefore, rectifying the tracks in the $g-T_{\rm eff}$ diagram,
i.e., inverting the gravity and multiplying by $\Te^4$, does not only turn the tracks
horizontal. As the luminosities of massive stars vary very little during
their evolution, this also allows the Y-axis of the sHR diagram to be much more stretched,
which makes the tracks of the most massive stars more distinguishable. 
An example of this is provided by Markova et al. (2014; their Fig.~6).
That this possibility has its limits when high- and low-mass
stars are shown together is demonstrated in Fig.\ref{fig_all}. 
We thus note here that the use of the effective gravity in Sect.~5 stretches the sHR diagram
close to the Eddington limit even more.

We note again that spectroscopic determinations of
$\log {\mathscr L \over {\mathscr L}_{\odot}}$
can be made with a precision of about 0.05\,dex, 
whereas the uncertainties of $\log g$ are a factor of two to
three larger (see Section\,2). This makes it difficult to determine masses from 
the $g-T_{\rm eff}$ diagram in Fig.\,1 for masses
larger than 30$\mso$, but allows Fig.\,2 to be used as a diagnostic tool for stellar masses.

Furthermore, it is interesting to compare the tracks of the stars in the three diagrams in
Figs.\ref{fig_grid} and \ref{fig_KiBo} that lose so much mass that their surface temperatures, after reaching a minimum
value, become hotter again. In the HR diagram, as the luminosities of these stars remain
almost constant during this evolution, it will thus be difficult
to distinguish whether an observed star is on the redward or on the blueward part of the track.
While it remains hidden in the HR diagram,
the sHR diagram reveals nicely that the mass loss drives these stars towards the Eddington limit.
As a consequence, the evolutionary state of observed stars will be much clearer in the sHR diagram.

\section{Stars with changing mass and helium abundance}
\label{sec_he}

\begin{figure*}[]
    \centering
     \includegraphics[angle=-90,width=8.5cm]{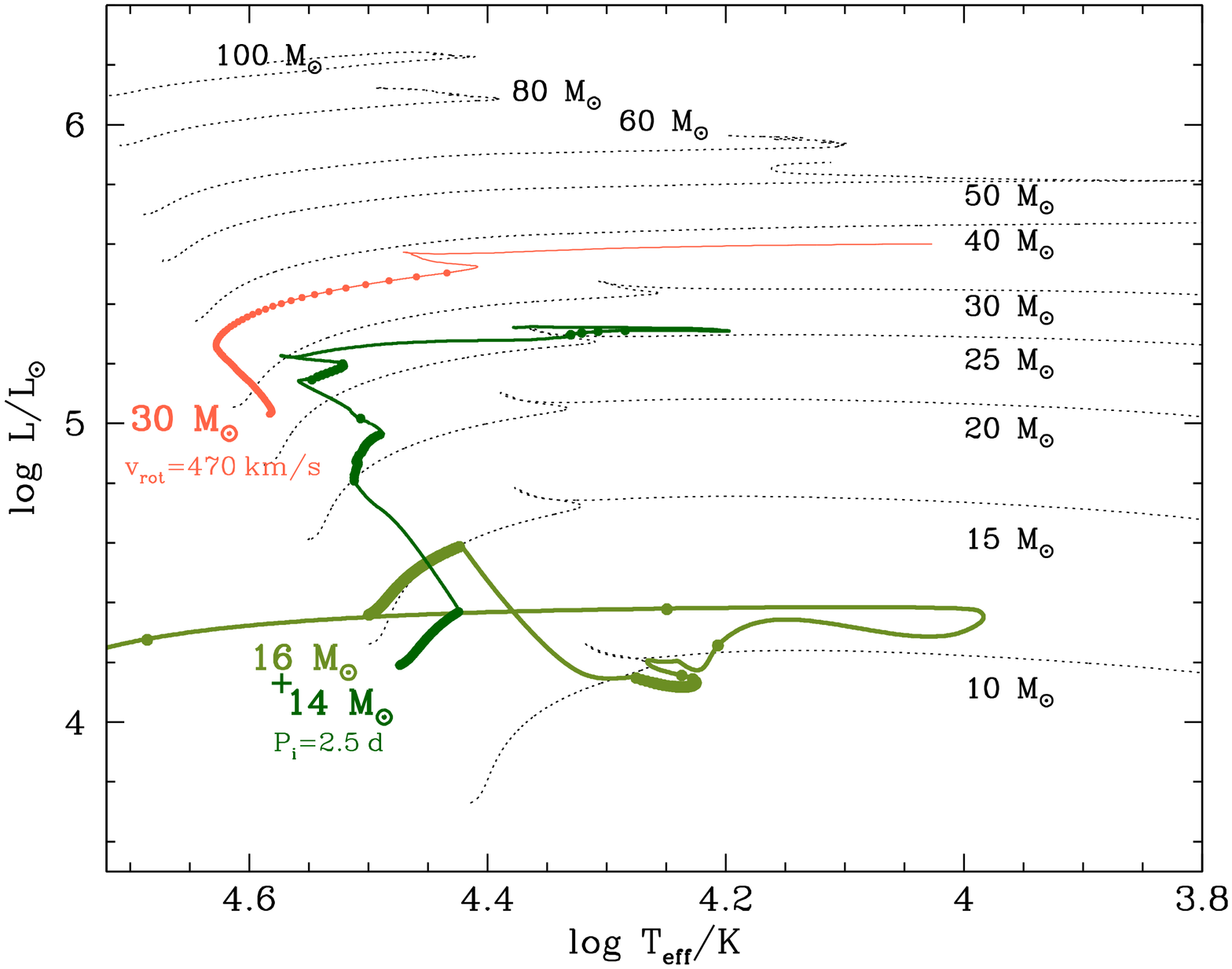} \includegraphics[angle=-90,width=8.5cm]{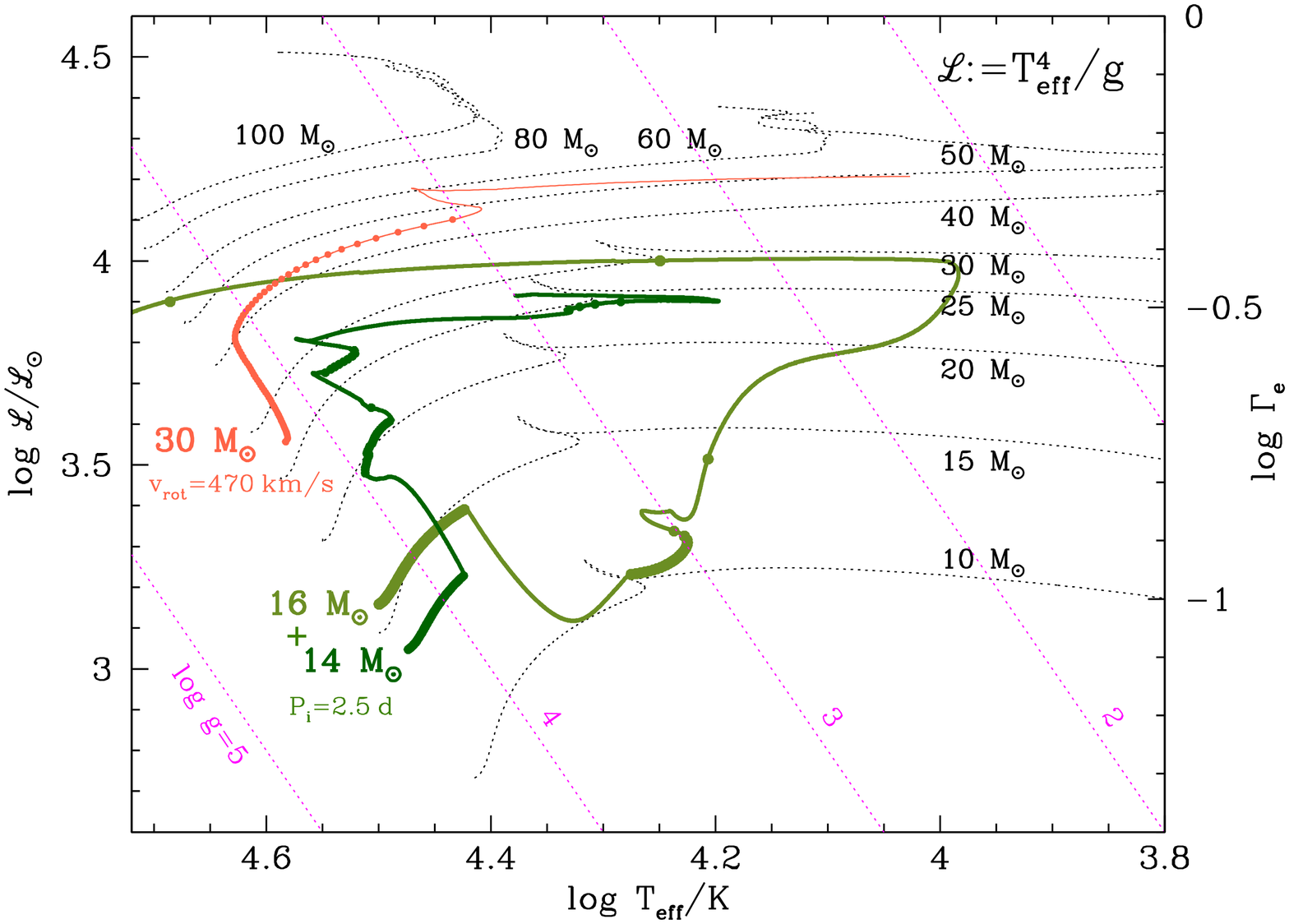}
    \caption{Evolutionary tracks of a rapidly rotating 30$\mso$ star, and of both components
of a 16$\mso$+14$\mso$ close binary, in the HR diagram (left) and in the sHR diagram (right).
Dots are placed on the tracks such that the stars spend $10^5\,$yr between two dots.
For comparison, the same tracks shown in Figs.~1 and~2 are included (black dotted lines).
The sHR diagram contains lines of constant gravity (dotted straight lines), and its
right ordinate scale is only valid for an unchanged surface helium abundance
of $Y\simeq 0.26$.}
    \label{fig_7}
\end{figure*}

For stars that evolve as ordinary single stars, one may use either the sHR diagram or,
for known distances, the HR diagram
to derive their properties, including their masses, in comparison to stellar evolution models
--- assuming here that models in particular for core hydrogen burning are trustworthy enough to allow this.
However, it is predicted that a certain fraction of stars does evolve according to unusual evolutionary paths,
in particular the close binary stars, which undergo mass transfer. Moreover, at high mass
there is also the possibility of chemically homogeneously evolving stars as a consequence of
rapid rotation (Brott et al. 2011). It is interesting to consider both types of evolution in the
HR and sHR diagram.
 
Both situations can lead to stars that are overluminous, i.e. stars that have a luminosity
larger than the luminosity of a single star of comparable mass and evolutionary state. 
Kippenhahn \& Weigert (1990) showed that main sequence stars are expected to obey the relation 
$L\sim M^{\alpha} \mu^{\beta}$, where $\mu$ is the average mean molecular weight and $\beta>1$.
An overluminosity is thus related to a larger mean molecular weight in a star than expected for
an ordinary star (Langer 1992). In the HR diagram, the ordinate ($L$) is increased by a factor of
$(1+\delta\mu)^{\beta}$, where $\delta\mu$ measures the excess in average mean molecular weight in our
overluminous star. Interestingly, in the sHR diagram, where the ordinate is proportional to
$L/M$ (independent of the helium abundance), the ordinate value of our star is increased by the same
factor over that of an ordinary star of the same mass. Because the ordinate in the HR diagram
depends more strongly on the mass ($L\sim M^{\alpha}$) than the ordinate in the
sHR diagram ($L/M\sim M^{\alpha -1}$), we conclude that the upward shift of our overluminous star
with respect to stellar evolution tracks for ordinary stars leads to higher apparent masses 
(i.e. masses determined by comparison to these tracks) in the sHR diagram than in the HR diagram.  

We can see this at the example of the evolutionary track of an extremely rapidly rotating 
30$\mso$ star (Brott et al. 2011) in the HR and sHR diagrams, in comparison to the tracks of
slowly rotating stars in Fig.~\ref{fig_7}. 
We note that in practice the determination of the surface gravity of a rapidly rotating star
may require a centrifugal correction (Herrero et al. 1992), and its inclination dependence may 
introduce additional uncertainties (Maeder 1999, Townsend et al. 2004). We do not consider this here,
but focus on the effect of the helium enrichment produced by the rapid rotation, which remains once
the star has spun down. The considered model evolves quasi-chemically homogeneously
until a helium mass fraction of about $Y=0.47$ is achieved, at which time it has spun down
sufficiently to switch to ordinary evolution. In both diagrams, the chemically homogeneous
part of the track covers a range of the ordinate value of $\sim 0.3\,$dex. Whereas in the HR diagram
this leads to an apparent mass of $M_{\rm HRD} \simeq 37\mso$, in the sHR diagram it leads to 
$M_{\rm sHRD} \simeq 42\mso$.
Because of stellar wind mass loss, the true mass of the model is slightly below 30$\mso$.

We see the same behavior in mass donors of interacting close binary models. 
This is demonstrated by the evolutionary 
track of the mass donor of a close binary model with an initial period of 2.5\,d and with initial masses
of 16$\mso$ and 14$\mso$ (system~42 in Wellstein et al. 2001) in the HR and sHR diagram
(Fig.~\ref{fig_7}). Its absolute and relative change in mass
is similar to that of the mass gainer (the binary evolution model is almost conservative). 
In the first mass transfer event, the mass donor loses about 9$\mso$.
Its core mass becomes much larger than that of a single stars of 7$\mso$, which corresponds
to its actual mass.
This increases its average mean molecular weight to values which ordinary single stars could not
achieve. An apparent mass of $M_{\rm HRD} \simeq 9.5\mso$ can be read off the HR diagram
(from the thick part of the track at $\log T_{\rm eff} \simeq 4.25$), while in the
sHR diagram an apparent mass of about $M_{\rm sHRD} \simeq 11\mso$ can be determined.

Figure~\ref{fig_7} shows that later, after the star returns from its minimum effective temperature,
the effect becomes even more drastic.
The mass donor is by then reduced to a total mass of 2.5$\mso$. Compared to an ordinary 2.5$\mso$ star,
its luminosity is increased by $\sim 2.4\,$dex, leading to an apparent mass of 
$M_{\rm HRD} \simeq 12\mso$ (again at 
$\log T_{\rm eff} \simeq 4.25$) in the HR diagram. A shift by $2.4\,$dex in the sHR diagram
leads a 2.5$\mso$ star from $\log \left( (L/\lso)/(M/\mso)\right) \simeq 1.6$ 
to $\log \left( (L/\lso)/(M/\mso)\right) \simeq 4$,
or an apparent mass of $M_{\rm sHRD} \simeq 30\mso$. We note that similarly drastic differences in the
apparent mass derived from the HR diagram from that derived from the sHR diagram can be expected
for post-AGB stars and post-red supergiant WNL stars. 

For the mass gainer in interacting binaries, there may
be almost no such effect. After the mass transfer, these stars may in many cases (though not 
always; see Langer 2012) rejuvenate and show global parameters ($R, L, T, g$) that 
are very similar to those of a single star of the 
same mass (except that they will appear younger than they are). This is shown by the evolutionary 
track of the mass gainer of the mentioned binary evolution model in Fig.~\ref{fig_7}.                    
The star with an initial mass of 14$\mso$ ends up with $\sim 26\mso$, and in both diagrams
it is located slightly above the 25$\mso$ single star track; i.e., among single stars, the 
location of the mass gainer in both diagrams would not be conspicuous. It would require a 
surface abundance analysis to show its enhanced nitrogen surface abundance and potentially fast
rotation to identify it as a binary product (Langer 2012), or its identification as a blue straggler
in a star cluster (Schneider et al. 2014). 

We note that overluminous stars may or may not have an increased helium surface abundance
(Langer 1992). While in practice, the determination of the effective temperature through model
atmosphere calculations may go along with the determination of the surface helium abundance,
the knowledge of the latter is not required in order to identify the position of the star in
either the HR or the sHR diagram. The surface helium abundance is only necessary for being able
to read off the Eddington factor from the right-side ordinate of the sHR diagram in Fig.~\ref{fig_7}.
In our figure, Eddington factors are given for a normal helium surface mass fraction ($Y_0 = 0.26$).
For other values, it can easily be adjusted according to  
\begin{equation}
\Gamma_{\rm e} = {\sigma\over c} {\mathscr L} \left( \sigma_{\rm e} \left(2-Y\right) \right)
\end{equation}
or
\begin{equation}
\Gamma_{\rm e}(Y) =  \Gamma_{\rm e}(Y_0) {2-Y\over 2-Y_0} .
\end{equation}
Compared to $Y_0 = 0.26$, the Eddington factor for helium-rich atmospheres 
can be reduced by up to a factor of 0.57, or 0.24\,dex. 

For stars with known distance, for which in addition to $g$ and $T_{\rm eff}$, $L$
can also be derived, the HR and the sHR diagram can be used together to identify 
overluminous stars. The actual mass follows directly from
the ordinate values of both diagrams as
\begin{equation}
\log {M\over \mso} = \log {L\over \lso} - log {{\mathscr L}\over {\mathscr L}_{\odot}},
\label{eq_mass}
\end{equation}
which corresponds to the spectroscopic mass, as it is effectively derived from
the spectroscopic gravity, effective temperature, and luminosity.
Only a reliable mass-luminosity relation is then required, or corresponding stellar
evolutionary tracks for ordinary single stars, to check whether the differences between
the observationally determined ordinate values and the ordinate values of an ordinary star with
the mass determined from Eq.\,\ref{eq_mass} are equal in both diagrams. 
If they are, then the true actual mass can be consistently determined from both diagrams.
Instead of using Eq.\,\ref{eq_mass}, it can also be determined from the apparent masses 
determined from the evolutionary tracks in both diagrams as
\begin{equation}
M={M_{\rm HRD}^{\alpha}\over M_{\rm sHRD}^{\alpha-1}},
\label{eq_mass1}
\end{equation}
where $\alpha$ is the exponent of the mass luminosity relation (e.g., Graefener et al. 2011).
We provide an example in Sect.\,\ref{sec_m81}. 

\section{Near the Eddington limit}
\begin{figure*}[]
    \centering
     \includegraphics[angle=-90,width=12.5cm]{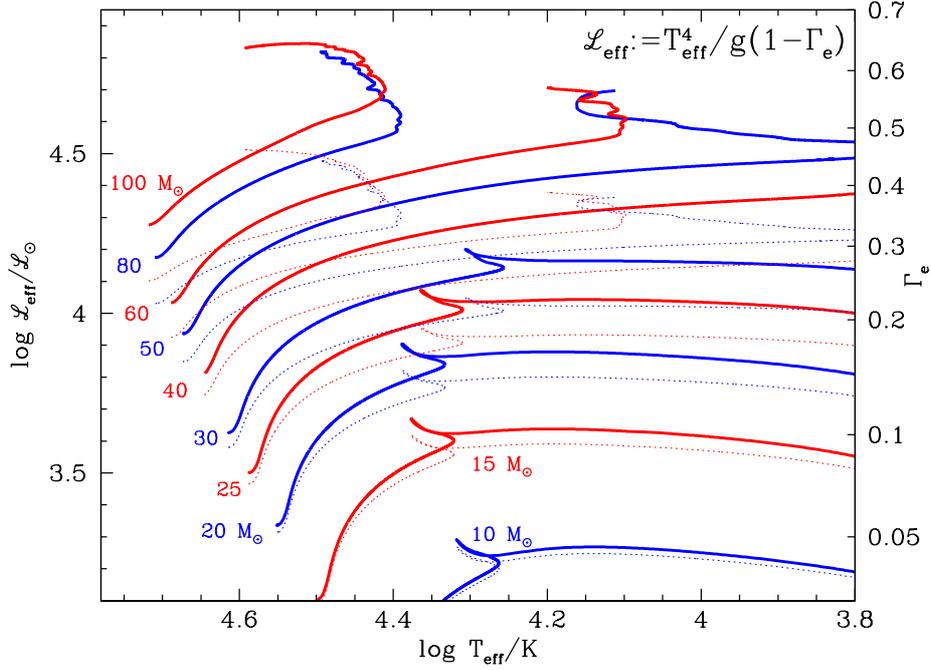}
    \caption{Evolutionary tracks for the same stellar evolution models shown in Fig.~2, 
here plotted in the effective sHR diagram (solid lines), for which the Eddington-factor can be read off 
from the ordinate on the right. For comparison, the tracks from Fig.~2 are also copied into this diagram
(dotted lines). The right ordinate scale is not valid for them.
The labels give the initial masses for the tracks drawn as solid lines.
}   
    \label{fig_eff}
\end{figure*}

In stars with a high luminosity-to-mass ratio, the radiation pressure may dominate the atmospheric structure.
Considering the equation of hydrostatic equilibrium in the form
\begin{equation}
{1\over \rho}{dP_{\rm rad}\over dr} + {1\over \rho}{dP_{\rm gas}\over dr} = {G M\over R^2} = g
\end{equation}
and replacing the first term by the photon momentum flux ${\kappa\over c} F_{\rm rad}$ at the stellar surface,
we can define as effective gravity the acceleration which opposes the gas pressure gradient term as
\begin{equation}
g_{\rm eff} = {G M\over R^2} - {\kappa\over c} F_{\rm rad} ,
\end{equation}
which can be written as $g_{\rm eff} = g (1-\Gamma)$.
In order to define an effective gravity that does not vary near the photosphere and in the wind acceleration zone,
we only consider the electron scattering opacity here and approximate
\begin{equation}
g_{\rm eff} \simeq g (1-\Gamma_{\rm e}) .
\label{eq_geff}
\end{equation}


Because of the high luminosity and the reduced surface gravity, stars near their Eddington limit tend to have strong
stellar winds. As a consequence, the stellar spectrum may be dominated by emission lines, and the ordinary gravity
may be hard to determine. However, model atmosphere calculations which include partly optically thick
outflows allow --- for stars with a known distance --- the stellar temperature, luminosity, and
radius to be determined (Hamann et al. 2006, Crowther 2007, Martins et al. 2008). From the widths of the emission lines, it is also
possible to derive the terminal wind speed, $\varv_{\infty}$. While not established for optically thick winds, 
optically thin winds show a constant ratio of the effective escape velocity from the stellar surface, 
$\varv_{\rm esc,eff}= \sqrt{2 G M (1-\Ge)/R}$, to the
terminal wind velocity over wide ranges in effective temperature (Abbott 1978, Kudritzki et al. 1992, Kudritzki \& Puls, 2000). 
Thus, assuming a relation of the form
\begin{equation}
\varv_{\infty} = r \varv_{\rm esc,eff},
\end{equation} 
where $r$ is assumed constant, would allow the calculation of the effective gravity from
\begin{equation}
g_{\rm eff} = {\varv_{\rm esc,eff}^2 \over 2 R}.
\end{equation}

It may thus be useful to consider a sHR diagram for massive stars where gravity is replaced by the effective 
gravity. We define 
\begin{equation}
{\mathscr L}_{\rm eff} = {\Te^4\over g(1-\Ge)} ,
\end{equation}
and consider an ``effective sHR diagram'' where we plot $\log {\mathscr L}_{\rm eff} / {\mathscr L}_{\odot}$
versus stellar effective temperature. Here, we approximate ${\mathscr L}_{\odot, \rm eff}$ by
${\mathscr L}_{\odot}$, as both quantities differ only by the factor $1/(1-\Gamma_{\odot})$,
where the solar Eddington factor, assuming electron scattering opacity (since we aim at massive stars),
is $\Gamma_{\odot}\simeq 2.6\, 10^{-5}$.

It is straightforward to plot evolutionary tracks in the effective sHR diagram, which is shown in 
Fig.~\ref{fig_eff}. We see that only above $\sim 15\mso$, the tracks deviate significantly from
those in the original sHR diagram. For more massive stars, the difference can be quite dramatic,
as seen for example from the tracks at $100\mso$. 

The topological character of the effective sHR diagram is different from that of the original
sHR diagram, as it no longer has a strict upper limit. However, instead of ${\mathscr L} = (c/\sigma \kappa_{\rm e})\Ge$,
we now have
\begin{equation}
{\mathscr L}_{\rm eff} =  {c \over \sigma \kappa_{\rm e}} {\Ge \over 1-\Ge} .
\end{equation}
While this means that we can still read the Eddington factor of stars directly off the effective sHR diagram
(Fig.~\ref{fig_eff}, right ordinate), the scale in $\log \Ge$ is no longer linear, but
instead it is 
\begin{equation}
\Ge = {1\over 10^{-\left(\log \left({\mathscr L}_{\rm eff}/{\mathscr L}_{\odot}\right) + \log \Gamma_{\odot}\right)} +1}.
\end{equation}

Furthermore, we see that ${\mathscr L}_{\rm eff} \rightarrow \infty$ for $\Ge \rightarrow 1$. 
Consequently, the effective sHR diagram conveniently stretches vertically for high $\Ge$,
in contrast to the ordinary sHR diagram. 
In practice the openness of the effective sHR diagram may not matter. It has been shown 
for massive star models of Milky Way and LMC metallicity that a value of $\Ge \simeq 0.7$
is not expected to be exceeded (Yusof et al. 2013, K\"ohler et al. 2014) --- because at 
roughly this value, massive stars reach their Eddington limit when the complete opacity is considered. 
Therefore, although this does not form a strict limit,
stars of the considered metallicity are not expected to exceed values of
${\mathscr L}_{\rm eff} \simeq 5$ (see Fig.\ref{fig_eff}). 
However, for stars of much lower metallicity, much higher values of ${\mathscr L}_{\rm eff}$ might be possible.

\section{The sHR diagram of low- and intermediate-mass stars}

\begin{figure}[]
    \centering
     \includegraphics[angle=-90,width=8.5cm]{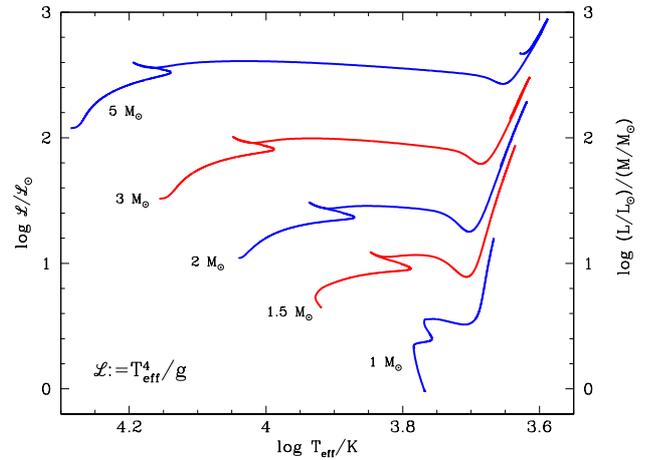}
   \caption{sHR diagram for low- and intermediate-mass stars.
   Because ${\mathscr L}\sim L/M$, one can use the spectroscopically
determined effective temperature and gravity to determine the
stellar mass-to-luminosity ratio (right Y-axis).
}
    \label{fig_low}
\end{figure}

In Fig.~\ref{fig_low}, we show the tracks of stars from $1\mso$ to $5\mso$ in the sHR diagram.
For stars in this mass range, the Eddington factor is small, and to be able to read it off the
sHR diagram may not be very relevant. Furthermore, except for the very hottest of these stars, the
true electron-scattering Eddington-factor is smaller than the indicated values, because the electron fraction
is reduced as a consequence of recombination of helium and hydrogen ions.
\begin{figure}[]
    \centering
     \includegraphics[angle=-90,width=8.5cm]{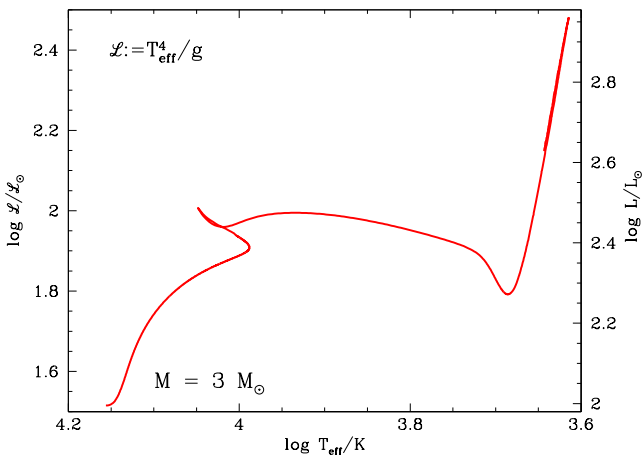}
   \caption{sHR diagram showing the evolutionary track of a
3$\mso$ star. This track is identical to a track in the HR diagram, and
the stellar luminosity is given by the alternative Y-axis on the right side of the diagram.
}
    \label{fig_3}
\end{figure}
However, independent of surface temperature and composition, it remains true that
${\mathscr L} \sim L/M$ (see Eq.~\ref{eq_lm}). Consequently, we have
\begin{equation}
\log {\mathscr L} / {\mathscr L}_{\odot} = \log\left(L/\lso\over M/\mso\right)
\end{equation}
(Fig.~\ref{fig_low}).

In the considered mass range, the sHR diagram has another advantage. At least until
very late in their evolution, these stars lose practically no mass, and so we have for a star, or evolutionary track,
of a given mass that
\begin{equation}
{\mathscr L} = {L \over k}  ,
\end{equation}
where the constant $k$ is $k=4\pi\sigma M$.
This is demonstrated in Fig.~\ref{fig_3}, which shows the evolution of a $3\mso$ star in the sHR diagram.
We note that the Y-axis to the right gives directly the stellar luminosity.

\section{Blue supergiants in the spiral galaxy M81: an extragalactic application}
\label{sec_m81}

\begin{figure}[]
    \centering
     \includegraphics[angle=90,width=8.5cm]{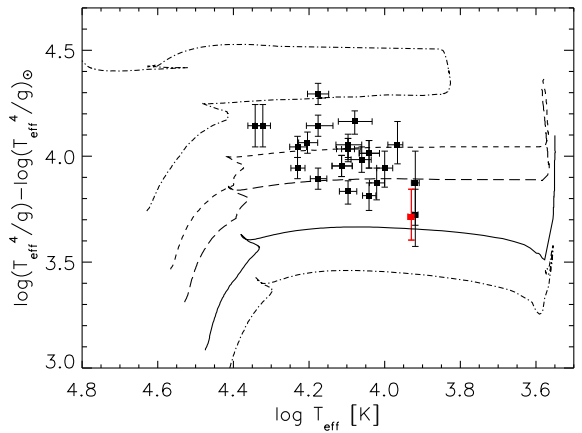}
     \includegraphics[angle=90,width=8.5cm]{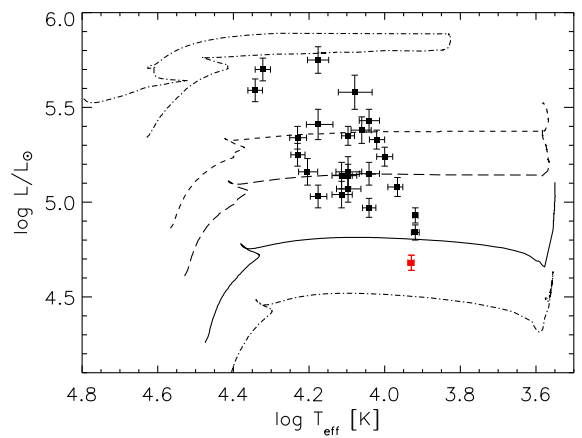}
   \caption{Blue supergiants in the spiral galaxy M81 at 3.47 Mpc distance. Top: sHR diagram; bottom: classical HR diagram. 
Spectroscopic data from Kudritzki et al. (2012). Evolutionary tracks  from Meynet and Maeder (2003) 
for Milky Way metallicity and including the effects of rotational mixing are shown 
(in increasing luminosity) for 12, 15, 20, 25, and 40 solar masses, respectively. 
The blue supergiant plotted in red is discussed in the text.}
    \label{fig_M81}
\end{figure}

Kudritzki et al. (2012) have recently carried out a quantitative spectroscopic study of blue supergiant stars 
in the spiral galaxy M81 with the goal of determining stellar effective temperatures, gravities, metallicities, 
and a new distance using the flux-weighted gravity--luminosity relationship. Figure\,\ref{fig_M81} shows the sHR diagram 
obtained from their results (we have omitted their object Z15, since its gravity is highly uncertain as 
discussed in their paper). The comparison with evolutionary tracks  nicely reveals the evolutionary status of 
these objects and allows to read their masses and ages off the sHR diagram without assuming a distance to the galaxy. 
As can be seen from  Fig.\,\ref{fig_M81}, the supergiants investigated are objects between 15$\mso$ and 40$\mso$, which have 
left the main sequence and are evolving at almost constant luminosity towards the red supergiant stage. 
For most objects the error bars are small enough to distinguish between the masses of the individual evolutionary tracks plotted. 
This is not possible for most of the objects when plotted in the 
corresponding $g-T_{\rm eff}$ diagram (shown in Fig.\,13 of Kudritzki et al.) 
mostly because the determination of flux-weighted gravity is more accurate than the determination of gravity 
(see Kudritzki et al., 2008, for a detailed discussion).

Since the distance to M81 is well determined (d = 3.47 $\pm$ 0.16 Mpc), we can compare the information 
contained in the sHR diagram with the one from the classical HR diagram, which is also displayed in Fig.\,\ref{fig_M81}. 
Generally, the conclusions with respect to stellar mass obtained from the two diagrams are consistent 
within the error bars. However, there are also discrepancies. The most striking example is the lowest luminosity 
object in the sample shown in red in both diagrams (object Z7 in Kudritzki et al.). While the HR diagram indicates 
a mass clearly below 15 M$_{\odot}$, the sHR diagram hints at a mass above this value. The reason for this discrepancy 
is that the spectroscopic mass of this object as derived from Eq.\,\ref{eq_mass} 
is only 9.2 M$_{\odot}$, whereas the mass one would assign to the object from its luminosity and based 
on the evolutionary tracks shown is 12.8  M$_{\odot}$ (see Table\,3 of Kudritzki et al.). 
Compared to a 9.2$\mso$ track, its luminosity is shifted by $\sim 0.55\,$dex in the HR diagram. 
Applying the test derived in Sect.\,\ref{sec_he}, we consider the shift of our star with respect to a 9.2$\mso$ 
track in the sHR diagram, and find it is practically the same as that in the HR diagram. 
We can thus confidently conclude that the observed star most likely has a mass of about 9$\mso$, 
and that it did not follow the ordinary single-star evolution, but  
is overluminous with respect to its mass.
As discussed by Kudritzki et al., 
mass discrepancies of this kind, while not frequent, have been encountered in many extragalactic blue supergiant 
studies and may indicate an additional mass-loss process not accounted for in the single-star evolutionary tracks,
or for an unusual mixing process inside the star.

\section {Concluding remarks}

We have shown above that the sHR diagram may be a useful tool for
deriving physical properties of observed stars, or for testing stellar evolution models.
The underlying reason is that when effective temperature and surface gravity are
spectroscopically determined, this provides a distance-independent measure of the
luminosity-to-mass ratio of the investigated star. The L/M-ratio is useful
to know in itself --- e.g., to determine the mass of a star cluster or a galaxy.
On the other hand, many stars, particularly low- and intermediate-mass stars, 
evolve at roughly constant mass, such that the
L/M-ratio remains proportional to the stellar luminosity. 
For high-mass stars, on the other hand, the L/M-ratio is proportional to their Eddington factor,
which is essential for their stability and wind properties.  

We have also demonstrated that for stars which are very close to their Eddington limit,
for which one can not determine the surface gravity spectroscopically because of their strong 
and partly optically thick stellar winds, it may be possible to consider a 
derivate of the sHR diagram where the gravity is exchanged with the effective gravity.
While it is still a challenge to derive the effective gravity observationally,
this may soon become possible with a better understanding of optically thick stellar winds.

In summary, while we believe that the original HR diagram, and the related color-magnitude diagram, 
will remain essential in stellar astronomy, 
the sHR diagram provides an additional tool for analysis that has not yet 
been widely explored and which may have the potential to supersede
the $g-T_{\rm eff}$ diagram in its original form,
as it appears to be more convenient and brings additional physical insight at the same time. 

\begin{acknowledgements} 
We thank our referee, Andre Maeder, for valuable remarks and comments.
NL is grateful to Norberto Castro, Luca Fossati, and Herbert Lau for inspiring discussions.
RPK acknowledges support from the National Science Foundation under grant AST-1008798 and the hospitality of 
the University Observatory Munich, where this work was carried out. 
\end{acknowledgements}

\bibliographystyle{aa}

\end{document}